\input ./style/arxiv-general.cfg
\documentclass[aoas,MSNbibl,nameyear,dvips]{arximspdf}
\makeatletter
   \@ifpackageloaded{graphicx}{}{\usepackage{graphicx}}
\makeatother

\doi{10.1214/15-AOAS816}
\volume{9}
\issue{2}
\pubyear{2015}
\firstpage{777}
\lastpage{800}
\docsubty{FLA}

\makeatletter
\newcommand{\lleft}{\left}
\newcommand{\rright}{\right}
\makeatother

\begin{document}
\begin{frontmatter}

\title{Covariance pattern mixture models for the analysis of
multivariate heterogeneous longitudinal data\thanksref{T1}}
\runtitle{CPMM for multivariate longitudinal data}
\thankstext{T1}{The HRS Diabetes Study is sponsored by the National Institute on Aging (Grant
Number NIA U01AG009740) and was conducted by the University of Michigan.}

\begin{aug}
\author[A]{\fnms{Laura}~\snm{Anderlucci}\ead[label=e1]{laura.anderlucci@unibo.it}}
\and
\author[A]{\fnms{Cinzia}~\snm{Viroli}\corref{}\ead[label=e2]{cinzia.viroli@unibo.it}}
\runauthor{L. Anderlucci and C. Viroli}
\affiliation{University of Bologna}
\address[A]{Department of Statistical Sciences\\
University of Bologna\\
via Belle Arti 41\\
Bologna 40126\\
Italy\\
\printead{e1}\\
\phantom{E-mail:\ }\printead*{e2}}
\end{aug}

%
\received{\smonth{1} \syear{2015}}
%
\revised{\smonth{2} \syear{2015}}

%
\begin{abstract}
We propose a novel approach for modeling multivariate longitudinal
data in the presence of unobserved heterogeneity for the analysis
of the Health and Retirement Study (HRS) data. Our proposal can be
cast within the framework of linear mixed models with discrete
individual random intercepts; however, differently from the standard
formulation, the proposed Covariance Pattern Mixture Model (CPMM)
does not require the usual local independence assumption.
The model is thus able to simultaneously model the heterogeneity,
the association among the responses and the temporal dependence
structure.

We focus on the investigation of temporal patterns related to the
cognitive functioning in retired American respondents. In particular,
we aim to
understand whether it can be affected by some individual
socio-economical characteristics and whether it is possible
to identify some homogenous groups of respondents that share a similar
cognitive profile. An accurate description of the detected groups
allows government policy interventions to be opportunely addressed.

Results identify three homogenous clusters of individuals with
specific cognitive functioning, consistent with the class
conditional distribution of the covariates. The flexibility of
CPMM allows for a different contribution of each regressor on the
responses according to group membership. In so doing, the
identified groups receive a global and accurate phenomenological
characterization.
\end{abstract}

%
\end{frontmatter}

\section{Introduction}\label{sec1}

The Health and Retirement Study (HRS) is conducted by the University of
Michigan every two years
[\citet{Juster1995}]. This panel study surveys a representative
sample of more than 26{,}000 Americans with 65 years and older, with
the aim of exploring the social, economic and health changes of
the respondents through an extensive questionnaire. It is a
multivariate longitudinal study where multiple responses on the same
individual are measured over a set of different occasions or times and,
as such, it has a three-way structure (see the next section for a
detailed description of the data).

One important goal of the study is the investigation of the cognitive
functioning of
the respondents in relation to the time and to potential socio-economic
covariates, so that government policy interventions could be addressed.

The cognitive functioning of an individual is a complex concept and it
is measured through several items of the questionnaire in the different
times. The association between the repeated measurements in a given
occasion and the temporal evolution of the cognitive functioning of the
individuals are two important aspects that a flexible model should be
able to describe.
A further issue that should be accounted for is the unobservable
heterogeneity between subjects that may not be explained by the
covariates. For instance, participants of the HRS study could
potentially have some cognitive impairments or dementia with a
different temporal pattern of their cognitive functioning. Thus,
heterogeneous individuals may belong to latent groups or classes that
differ because they may exhibit different temporal patterns of their
cognitive functioning and different association among the responses
that define their cognitive status.

A variety of approaches for modeling multivariate
longitudinal data have been proposed in the statistical literature
in recent years. They can be disentangled into multivariate
longitudinal factor models and random effects models [see, for a
comprehensive review,
\citet{Verbeke2013} and \citet{Bandyopadhyay2011}].

In the former family of methods, it is assumed that one or more
underlying variables explain the association among the multiple
responses, thus
reducing the dimensionality problem. The approach can be cast
within the wide framework of the Structural Equation Modeling
(SEM). See, for example, \citet{Ferrer2003,Timmerman2003,Fiews2009} and \citet{Vasdekis2012}, among
others.

Random effects models or growth curve models assume that
repeated measurements of a particular response represent
realizations of a latent subject-specific evolution through the
inclusion of subject-specific parameters
[see \citet{Laird1982} and \citet{Reinsel1984}] that
typically have a continuous distribution.
These models belong to the class of generalized linear mixed models [see
\citet{Goldstein1995,Muthen2002,McCulloch2008} and \citet{Skrondal}].

All these methods are developed under the implicit assumption of
homogenous individuals over time.
In order to deal with heterogeneous observations, as it is in our case,
the simplest idea consists of the inclusion of individual-specific
random intercepts that have a discrete distribution. These models are
forms of latent class models [see \citet{Lazarsfeld} and
\citet{Vermunt}] and mixture models [\citet{Quandt1978,McLachlan2000,Fraley2002}]. In longitudinal
data analysis, the random intercepts are typically assumed to be
time-varying, {that is}, they are associated to latent temporal
trajectories via latent autoregressive models or, alternatively, latent
Markov models [\citet{Bartolucci}]. See \citet{Bartolucci2}
for a nice review and comparison between the two formulations.

The framework of discrete (time-constant or varying) random intercepts
for modeling heterogeneity includes mixture random effect models for
univariate longitudinal data [\citet{Verbeke1996}], recently
extended to deal with multivariate and mixed outcomes by \citet
{ProustLima2005} and \citet{ProustLima2012}, and growth mixture
models, where individuals are
grouped in classes having a specific growth structure variability
within them [see \citet{Muthen2007}].

In a model-based clustering perspective, \citet{Manrique}
introduced a clustering strategy based on a mixed membership framework
for analyzing
discrete multivariate longitudinal data. For continuous responses,
\citet{DelaCruz2008} proposed a mixture of hierarchical nonlinear
models for describing nonlinear relationships across
time. \citet{McNicholas2010} introduced a family of Gaussian mixture
models by parameterizing the class conditional covariance matrices
via a modified Cholesky decomposition, that
allows to interpret the observations as derived by a generalized
autoregressive process and to explicitly incorporate their
temporal correlation into the model. Both approaches focus on
model-based clustering of a single response measured on a set of
different occasions. Alternatively, \citet{Leiby} proposed a
multivariate growth curve mixture model that groups subjects on the
basis of multiple symptoms measured repeatedly over time. They
developed their approach by assuming a within-class latent factor
structure explaining the correlations among the responses.
Alternatively, nonparameric Bayesian approaches have been becoming
increasingly popular for modeling longitudinal data thanks to the
Dirichlet process prior that allows for an infinite dimensional number
of classes, thus capturing the heterogeneity in a very flexible way
[see, among others, \citet{Muller,Muller2,kleinman,Brown} and the references therein].

In this paper, we propose a model for multivariate longitudinal data
which is based on a mixture of latent generalized autoregressive
processes with order $m$ ($m<T$, where $T$ is the number of observed
time points). In our formulation the observable variables are not
required to be independent
given the latent states (\emph{local independence} assumption): in
fact, we account simultaneously for the association between the
responses and for the unobserved heterogeneity between subjects in the
dynamic observational process. To the best of our knowledge, the
classical approaches for the analysis of longitudinal data hardly
account simultaneously for the three goals of the analysis, which arose
from the three layers of the data structure: heterogeneous units,
correlated occasions and dependent variables.

In what follows, we will present our proposal in three gradual steps in
order to sequentially address the three issues, so as to finally define
the complete model we can refer to as the Covariance Pattern Mixture
Model (CPMM). Each component of the mixture corresponds to a state of a
discrete random intercept and identifies a group of individuals with
the same temporal profile and similar effect of the covariates. In this
perspective, the proposed model belongs to the class of mixtures of
regression models [\citet{Grun2007}]. As such, it can be also
viewed as an extension of the proposal of \citet{McNicholas2010}
in the multivariate context. The proposed approach will be illustrated
in Section~\ref{sec3}.

In order to make inference on the proposed model, we adopt
the matrix-normal distribution [\citet{Dutilleul1999}] for modeling
the density of the outcomes observed in the different times
conditionally to each class of observations. In so doing, we
assume equally spaced and balanced data across subjects and with regard
to the responses at each occasion. Each class-distribution is
characterized by the separability condition
of the total variability into two sources related
to the multiple attributes and to the temporal evolution via the
Kronecker product, in the same perspective of \citet{Naik2001}.
Although seemingly
complex, the model can be fitted using an
expectation--maximization (EM) algorithm. Compared to other methods for
the analysis of multivariate longitudinal data, the algorithm
convergence is pretty fast, despite the dimensionality of the problem.
The observed information matrix can be derived numerically and
exploited to obtain standard errors for the regression coefficient
estimates. Estimation details are presented in Section~\ref{sec4}. In the
supplementary material, a large simulation study is illustrated, aiming
at validating the proposed model in terms of robustness and accuracy.

The flexibility of the proposed model and the advantages with respect
to alternative proposals are illustrated through the application to the
longitudinal data on cognitive functioning of the HRS in Section~\ref{sec5}. A
discussion of the model results in relation to their political and
social implications is presented in Section~\ref{sec6}. Section~\ref{sec7} contains some
final remarks on the proposed approach.

\section{HRS data description}\label{sec2}
We consider data coming from the Health and Retirement Study
(HRS) started in 1992 and conducted by the University of Michigan (USA)
every two
years. It is a panel study that surveys a
representative sample of more than 26{,}000 Americans of age
65 years and older (\surl{http://hrsonline.isr.umich.edu/}) with the aim
of collecting information about income, work, assets, pension plans,
health insurance, disability, physical health and functioning,
cognitive functioning and health care expenditures. The HRS allows to
explore the health changes that individuals undergo toward the end
of their work lives and in the years that follow. This survey comprises
a more extensive study on Aging, Demographics and Memory (ADAMS) on a
wave of 856 subjects, selected from the total sample frame of
approximately 26{,}000 HRS individuals.

A description of the scientific, public policy and organizational
background of the study can be found in \citet{Juster1995},
whereas the details of the ADAMS sample design are described in
\citet{Heeringa2007}.

Many aspects have been investigated on this database so far.
\citet{Lang2005} linked the ADAMS dementia clinical assessment data to the
wealth of available longitudinal HRS data to study the onset of
Cognitive Impairment, Non Demented (CIND), as well as the risk factors,
prevalence, outcomes, and costs of CIND and dementia.

\citet{McArdle2007} used contemporary latent variable models to
organize information in terms of both cross-sectional and longitudinal
inferences about age and cognition, with the aim of better describing
age trends in cognition among older adults in the HRS study from 1992
to 2004.

Furthermore, \citet{Plassman2008} estimated the prevalence of
cognitive impairment without dementia in the United States and
determined longitudinal cognitive and mortality outcomes. In
\citet{Steffens2009} the national prevalence of depression,
stratified by age, race, sex and cognitive status, was estimated.
Logistic regression analyses were performed to examine the association
of depression and previously reported risk factors for the condition.

In our study, one of our aims is to investigate temporal patterns of the
cognitive functioning in order to understand whether it can be affected
by some individual characteristics and
whether it is possible to identify some homogeneous groups of
respondents that share a similar cognitive profile.

\begin{figure}[t]

\includegraphics{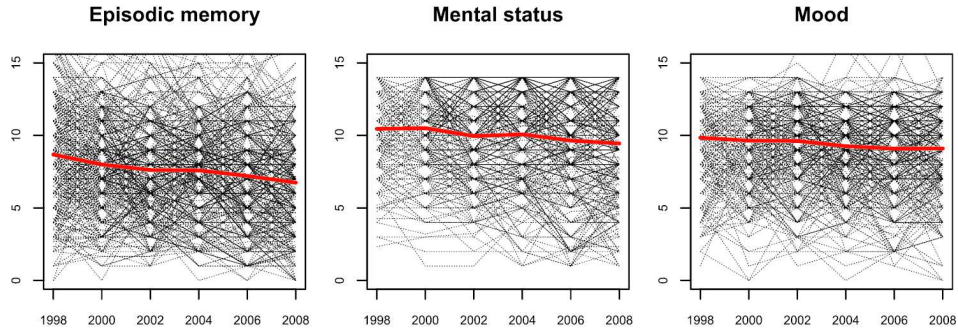}

\caption{Individual trajectories (dashed black lines) and mean profiles
(solid red lines) of the three responses during the 6 time
points.}\label{figmeanprofiles}
\end{figure}

In order to accomplish this objective, we consider the sample of 359
individuals among 856 subjects, for whom the information is complete
without missing in some entries in the years from 1998 to 2008.
This sample is a cohort followed in all waves without refreshment: the
same individuals were surveyed from 1998 to 2008 every two years, for a
total time span of 10 years and 6 time points (i.e., in 1998, 2000, 2002,
2004, 2006, 2008). Three responses
are investigated, namely, the ``episodic memory'' (EM), the ``mental
status'' (MS) and the ``mood'' (MO); they represent a summary of
several assessment questions. Their scores are positively related to
the performance of the individuals in the corresponding dimension. The
observed mean profile plots of the three
responses in Figure~\ref{figmeanprofiles} show different
patterns in times, suggesting the need of a proper model
able to account for its dynamic.

The cognitive functioning in a given time may indeed depend on its past
values measured in previous occasions.
The correlation across time points (see Table~\ref{tabcortemp})
suggests that there is a temporal association, since the values are
pretty high, particularly when considering consecutive or close moments.

\begin{table}[b]
\tablewidth=230pt
\caption{Temporal correlation matrix}\label{tabcortemp}
\begin{tabular*}{230pt}{@{\extracolsep{\fill}}lcccccc@{}}
\hline
& \textbf{1998} & \textbf{2000} & \textbf{2002} & \textbf{2004} & \textbf{2006} & \textbf{2008}
\\
\hline
1998 & 1.000 & & & & &\\
2000 & 0.638 &1.000 & & & & \\
2002 &0.638 &0.680 &1.000 & & & \\
2004 &0.603 &0.643 &0.708 & 1.000 & & \\
2006 &0.572 &0.622 &0.686 &0.715 & 1.000 & \\
2008 &0.564 &0.633 &0.665 &0.702 &0.733 &1.000\\
\hline
\end{tabular*}
\end{table}

Furthermore, Table~\ref{tabcorp} shows that an association among the
different aspects of the cognitive ability is present too: the
considered responses have a mild but significant correlation (all
$p$-values${}< 2.2\mathrm{e}{-}16$).

We also consider some other demographic and socioeconomic
information on the respondents that may have an effect on the
responses, in particular:
\begin{itemize}
\item \emph{gender}, coded as ``0'' if males and as ``1'' if females;
\item \emph{age}, taken as numeric;
\item \emph{level of education}, in terms of years of school;
\item \emph{health self-rating}, coded as ``1'' if considered
``excellent,'' ``2'' if ``very good,'' ``3'' if ``good,'' ``4'' if ``fair''
and ``5'' if ``poor.''
\end{itemize}

Table~\ref{tabdescrittive} shows some descriptive statistics of
the considered covariates. The majority of the respondents are
females (57.1\%), with an average age of 74.3 in 1998 (first time
point considered). The average number of years of education is
about 11, while the average rating of perceived health is about 2.8 in 1998.

\begin{table}[t]
\tablewidth=200pt
\caption{Response correlation matrix}\label{tabcorp}
\begin{tabular*}{200pt}{@{\extracolsep{\fill}}lccc@{}}
\hline
& \textbf{Episodic} & \textbf{Mental} & \\
& \textbf{memory} & \textbf{status} &  \textbf{Mood}\\
\hline
Episodic memory & 1.000 & & \\
Mental status & 0.520 & 1.000 & \\
Mood & 0.200 & 0.219 & 1.000\\
\hline
\end{tabular*}
\end{table}

\begin{table}[b]
\tablewidth=280pt
\caption{HRS data: Descriptive statistics of the covariates}
\label{tabdescrittive}
\begin{tabular*}{280pt}{@{\extracolsep{\fill}}lcccc@{}}
\hline
\textbf{Variable} & \textbf{Details} & \textbf{\%} & \textbf{Mean} & \textbf{Standard deviation} \\
\hline
Gender & Female & 57.1 & -- & -- \\
& Male & 42.9 & -- & -- \\
Age &(in 1998)& -- & 74.3 & 5.6 \\
Education & (in years) & -- & 11.1 & 6.2 \\
Health self-rating & (in 1998) & -- & \phantom{0}2.8 & 1.1 \\
\hline
\end{tabular*}
\end{table}

The aim is to fit a model that is able to capture the temporal
evolution of the cognitive functioning, to explain the association
among the responses and to simultaneously account for unobserved
heterogeneity among the units. The selected covariates may help in the
characterization of the phenomenon, so that ad hoc interventions to
take care of elderly people needs can\ be made.

\section{Model formulation}\label{sec3}

\subsection{Modeling the unobserved heterogeneity}\label{sec31}
We first consider the problem of modeling the unobserved heterogeneity
in the univariate context, where the number of responses, say $p$, is
$p=1$. The extension to $p>1$ will be developed in Section~\ref{sec33}. Suppose
we observe a continuous response on $n$ individuals and on each of
them, observations are taken over $T$ time points. We denote with
$y_{jt}$ the response for subject $j$ ($j=1,\ldots,n$) at occasion $t$
($t=1,\ldots,T$) and with $\mathbf{x}_{jt}$ the corresponding vector
of $q$ covariates. The simple linear regression model
\[
y_{jt}=\alpha+ \mathbf{x}_{jt}^\top\bolds{\theta}+
\varepsilon_{jt},
\]
with intercept $\alpha$, regression coefficients $\bolds{\theta}$
and Gaussian residuals $\varepsilon_{jt} \sim\phi(0,\sigma_\varepsilon^2)$,
could be extended to account for the unobserved heterogeneity by
including individual-specific random intercepts and (or) random slopes.
A variety of mixed models can be defined depending on whether
continuous or discrete random effects are considered [\citet
{Laird1982}]. One aim of the HRS data analysis is the identification of
groups of subjects with similar, say, cognitive functioning that could
potentially correspond to specific mental health conditions.
For this reason, we consider a discrete parameterization for the random
effects. Let $\alpha_j$ be the subject-specific random intercept that
may assume $k$ possible values, denoted by $\theta_{0i}$, with some
probabilities, say $\pi_i$, with $\sum_{i=1}^k \pi_i=1$ for $i=1,\ldots
,k$. This is equivalent to assuming the mixture model
%
\begin{equation}
\label{equaz1} f(y_{jt})=\sum_{i=1}^k
\pi_i \phi\bigl(\theta_{0i}+\mathbf{x}_{jt}^\top
\bolds{\theta}, \sigma_\varepsilon^2\bigr).
\end{equation}
A closer look to (\ref{equaz1}) shows that this formulation is barely
useful, unless we allow either regression coefficients $\bolds{\theta}$ or the variance $ \sigma_\varepsilon$ (or both) to be somehow
dependent on the state of $\alpha_j$, since otherwise the heterogeneity
structure could be hardly captured by the model. Thus, a general
formulation of a full heterogeneous model is
%
\begin{eqnarray}
\label{equaz2} && y_{jt}=\theta_{0i}+\mathbf{x}_{jt}^\top
\bolds{\theta}_i + \varepsilon_{ijt} \qquad \mbox{with
probability } \pi_i,
\end{eqnarray}
where $ \varepsilon_{ijt} \sim\phi(0,\sigma_i^2)$. Thus, we obtain
%
\begin{equation}
f(y_{jt})=\sum_{i=1}^k
\pi_i \phi\bigl(\theta_{0i}+\mathbf{x}_{jt}^\top
\bolds{\theta}_i, \sigma_i^2\bigr).
\end{equation}
This formulation is based on the assumption that, for every unit $j$,
the response at the different occasions is conditionally independent
given the covariates and the individual-specific intercept denoting the
group membership. This condition, well known as \textit{local
independence}, is quite restrictive in practice, since the temporal
observations could be highly correlated, especially with the most
recent past.

\subsection{Modeling correlated temporal data}\label{sec32}
The most common formulation for modeling the temporal correlation in
longitudinal data consists of introducing continuous time-varying
individual random effects that follow an autoregressive latent model of
order 1, AR(1) [\citet{Chi}] with correlation coefficient $\rho$:
\begin{eqnarray*}
&& y_{jt}=\alpha_{jt}+\mathbf{x}_{jt}^\top
\bolds{\theta}+ \varepsilon_{jt},
\end{eqnarray*}
with
\begin{eqnarray*}
\alpha_{j1} &=&u_{j1},
\\
\alpha_{jt} &=& \alpha_{j(t-1)} \rho+ u_{jt}
\sigma_u,\qquad t=2,\ldots,T
\end{eqnarray*}
and $u_{jt}\sim\phi(0,1)$.
The model could be extended to allow for random slopes besides the
random intercepts in a very parsimonious way [see \citet{Goldstein1995,McCulloch2008} and \citet{Skrondal}].

When $\alpha_j$ has a discrete formulation (such as in our case) the
temporal correlation can be modeled by assuming an autoregressive
process on the error term $ \varepsilon_{ijt}$. Here we assume that $
\varepsilon_{ijt}$ follows a latent Generalized AutoRegressive (GAR)
process of generic order $m$:
%
\begin{eqnarray}
\label{equaz4} &&\varepsilon_{ijt}=\sum_{s=1}^{\min(m,t-1)}(-
\rho_{it(t-s)})\varepsilon _{ij(t-s)} + u_{jt}
\sqrt{d_{it}},
\end{eqnarray}
where $d_{it}$ are time-varying constants representing the innovation
variances. In (\ref{equaz4}) the summation is empty and its value is
zero if the lower bound is greater than the upper
bound $\min(m,t-1)$. $m$ is the order of the process and it can range
in $\{0,1,\ldots,T-1\}$. The value $m=0$ means temporal independence,
$m=1$ denotes a generalized
autoregressive process of order 1, and so on, until the full model
with $m=T-1$ that corresponds to the less interesting situation
of not restricted temporal structure. Notice that when $d_{it}=d_{i}$
for all $t=1,\ldots,T$ the GAR coincides with the AutoRegressive (AR)
process of order $m$.

The model (\ref{equaz2}) without covariates extended with the GAR
structure in (\ref{equaz4}) has been proposed by \citet
{McNicholas2010} and applied to yeast sporulation time course data. The
authors developed a family of mixture models by observing that the
generalized autoregressive process in (\ref{equaz4}) is equivalent to
assuming a modified Cholesky decomposition of the $T$-dimensional
temporal covariance matrix, say $\Phi$.
The modified Cholesky decomposition [\citet{Newton1988,Pourahmadi1999}] establishes that a matrix $\Phi$ is
positive definite if and only if there exits a unique unit lower
triangular matrix $U$, with 1's as diagonal entries, and a unique
diagonal matrix $D$ such that $U\Phi U^\top=D$ or, equivalently,
$\Phi^{-1}=U^\top D^{-1} U$. More specifically, the matrix $U$
takes the form
\[
U=\lleft[ %
\matrix{ 1 & 0 & \cdots & \cdots& \cdots & \cdots & 0
\vspace*{2pt}
\cr
-\rho_{1,2} & 1 & 0 & \cdots& \cdots & \cdots & 0
\vspace*{2pt}
\cr
-\rho_{1,3} & -\rho_{2,3} & 1 & 0 & \cdots &
\cdots & 0 \vspace*{2pt}
\cr
\cdots & \cdots & \cdots & \ddots& 0 & \cdots & \cdots
\vspace*{2pt}
\cr
-\rho_{1,t}& -\rho_{2,t} & \cdots & \cdots& 1 &
\cdots & \cdots\vspace*{2pt}
\cr
\cdots & \cdots & \cdots & \cdots& \cdots &
\ddots & 0 \vspace*{2pt}
\cr
-\rho_{1T} & -\rho_{2,T} & \cdots &
\cdots& \cdots & -\rho_{T-1,T} & 1 } \rright],
\]
while $D$ is a $T \times T$ diagonal matrix with
positive entries $d_t$  ($t=1,\ldots,T$), that represent the innovation
variances.

Formulation (\ref{equaz2}) together with (\ref{equaz4}) is equivalent
to assuming the following mixture model for the $T$-dimensional vector
$\mathbf{y}_{j}$:
%
\begin{eqnarray}
\label{equaz5} && f(\mathbf{y}_{j})=\sum_{i=1}^k
\pi_i \phi\bigl(\bolds{\theta}_{0i} + X_{j}
\bolds{\theta}_i, \bigl(U_i^\top
D_i^{-1}U_i\bigr)^{-1}\bigr),
\end{eqnarray}
where $X_j$ is the matrix of $q$ covariates of dimension $T \times q$ and
$\bolds{\theta}_{0i}$ is the $T$-dimensional vector containing the
intercepts. To give more flexibility to the model, we allow for $T$
time-varying intercepts for each group so that $\bolds{\theta}
_{0i}=[\theta_{0i1},\ldots,\theta_{0iT}]$.

\subsection{Modeling multivariate longitudinal data: Covariance pattern
mixture models}\label{sec33}
When $p>1$, a common assumption for modeling multivariate longitudinal
data is the local independence, that is, the observed variables are
assumed to be mutually independent given the latent states. We do not
require the local independence between the responses, as we explicitly
model their association. This is achieved by extending the model in the
form of a matrix-variate regression model [\citet{Viroli2012}]
with a discrete random intercept in order to take into account the
correlations among the $p$ responses:
%
\begin{eqnarray}
\label{equaz6} && Y_{j}= \bolds{\theta}_{0i}
\mathbf{c}^\top + {X}_{j} \Theta_i +
E_{ij} \qquad \mbox{with probability } \pi_i,
\end{eqnarray}
where $Y_j$ is a matrix of continuous responses of dimension $T \times
p$, $X_j$ is the matrix of $q$ covariates of dimension $T \times q$,
$\mathbf{c}$ is a $p$-dimensional vector of ones, $\Theta_{i}$ is a
matrix of dimension $q \times p$ containing the regression coefficients
and $E_{ij}$ is a $T \times p$ matrix of error terms distributed
according to the matrix-normal distribution [\citet
{Dutilleul1999}]. This probabilistic model can be thought of as an
extension of the multivariate Gaussian distribution for modeling
continuous random matrices instead of the conventional vectors.
Let $\Phi$ be a $T \times T$ covariance matrix containing the variances
and covariances between
the $T$ times and $\Omega$, a $p \times p$ covariance matrix
containing the variance and covariances of the $p$ responses. The
matrices $\Phi$ and $\Omega$ are commonly referred to as the \emph
{between} and the \emph{within} covariance
matrices, respectively. The $T \times p$ matrix-normal
distribution is defined as
\begin{eqnarray*}
f(E|\Phi,\Omega)&=&(2\pi)^{-({Tp})/{2}}|\Phi|^{-{p}/{2}}|\Omega
|^{-{T}/{2}} \exp \bigl\{-\tfrac{1}{2}\operatorname{tr}
\Phi^{-1}E\Omega^{-1}E^\top \bigr\}
\end{eqnarray*}
or, in compact notation, $E \sim\phi^{(T \times p)}(0,\Phi,\Omega)$.

It is easy to show that a matrix-normal distribution has an
equivalent representation as a multivariate normal distribution of
dimension $T\times p$, with covariance matrix, say $\Sigma$,
separable in the form $\Sigma=\Phi\otimes\Omega$ (where
$\otimes$ is the Kronecker product). The separability condition
has the twofold advantage of allowing the modeling of the temporal
pattern of interest directly on the covariance matrix $\Phi$ and of
representing a
more parsimonious solution than that of the unrestricted $\Sigma$, with
a number of parameters equal to $p(p+1)/2 + T(T+1)/2$ instead of $pT(pT+1)/2$.
Moreover, notice that the restricted model under the local independence
assumption referred to the temporal observations (or to the responses)
can be obtained by taking $\Phi$ (or $\Omega$) equal to the identity matrix.

Let $M_{ij}$ be the systematic part of the model, that is,
$M_{ij}=\bolds{\theta}_{0i}\mathbf{c}^\top+ {X}_{j} \Theta
_i=\tilde{X}_j \tilde{\Theta}_i$, where $\tilde{X}_j$ is the matrix of
covariates of dimension $T \times(T+q)$; the sub-matrix corresponding
to the first $T$ columns is an identity matrix designed to incorporate
an intercept term for each time point and $\tilde{\Theta}_i$ is a
$(T+q)\times p$ matrix of regression parameters.

The model (\ref{equaz6}) can be rephrased as a mixture model of $k$
matrix-normal distributions of sizes $\pi_1,\ldots,\pi_k$, with mean
matrices $M_{ij}$, and two covariance matrices: $\Omega_i$ is the
response covariance matrix and $\Phi_i$ is a temporal covariance matrix
that can be decomposed according to the modified Cholesky
representation. More specifically, the density of the generic observed matrix
$Y_j$ is defined as
%
\begin{eqnarray}
\label{eqn1} && f(Y_j|\bolds{\pi},\bolds{\Theta})=\sum
_{i=1}^k\pi_i\phi^{(T
\times p)}
(Y_j;M_{ij},\Phi_i,\Omega_i ),
\end{eqnarray}
where $\Phi_i= (U_i^\top D_i^{-1}
U_i )^{-1}$ and $\bolds{\Theta}_i=\{\Theta_i,U_i,D_i,\Omega_i\}
$ with $i=1,\ldots,k$ collectively
denote the set of matrix normal parameters. The component
density in (\ref{eqn1}) is given by
\begin{eqnarray*}
&& \phi^{(T \times p)} (Y_j;M_{ij},\Phi_i,
\Omega_i )\\
&&\qquad=(2\pi)^{-({Tp})/{2}} |D_i|^{-{p}/{2}}
\times |\Omega_i|^{-{T}/{2}}
\\
&&\qquad\quad{}\times \exp \bigl\{ -\tfrac{1}{2} \operatorname{tr} \bigl(U_i^\top
D_i^{-1} U_i\bigr) (Y_j-\tilde
X_j \tilde\Theta_i)\Omega_i
^{-1}(Y_j-\tilde X_j \tilde
\Theta_i)^\top \bigr\}.
\end{eqnarray*}

If no restriction is imposed on the mixture component parameters, the
proposed mixture model is very flexible since classes can differ
with respect to specific temporal patterns and according to the
class conditional variability of the responses. However, the
number of parameters in the matrix-variate formulation could be high
with respect to the sample size. In
addition, in some applications it could be of interest to
investigate whether the potential groups of individuals vary with
respect to both a different temporal correlation and a specific
variable variation, or with respect to one of the two sources
only. By allowing some but not all of the matrices $\Omega_i$,
$U_i$ and $D_i$ to vary among clusters, a family of different
mixture models can be defined and explored.

\begin{table}[b]
\caption{Pattern covariance structures and number of parameters}\label{tabcovariancestructure}
\begin{tabular*}{\tablewidth}{@{\extracolsep{\fill}}llc@{}}
\hline
 &  & \textbf{No. of
covariance} \\
\textbf{Pattern}&\multicolumn{1}{c}{\textbf{Description}} & \textbf{parameters}\tabnoteref{t1} \\
\hline
Nontemporal & \\
\quad VVV & Heteroscedastic components & $k \frac{p(p+1)}{2}$ \\[3pt]
\quad EEV & Ellipsoidal, equal volume and equal shape & $p + k \frac
{p(p-1)}{2}$\\[3pt]
\quad EEE & Homoscedastic components & $\frac{p(p+1)}{2}$\\[2pt]
\quad III & Spherical components with unit volume & $0$\\
\quad VVI & Diagonal but varying variability components & $ kp$ \\
\quad EEI & Diagonal and homoscedastic components & $ p$\\
\quad VII & Spherical components with varying volume & $ k $\\
\quad EII & Spherical components without varying volume & $1$\\[3pt]
Temporal & \\
\quad  $\operatorname{GAR}(m)$ & Heteroscedastic components, $m = 0,1,\ldots,T-1$ & $kT+k\phi$\\
\quad  $\operatorname{GARI}(m)$ & GAR${}+{}$isotropic for $D$ &$ k+k\phi$\\
\quad  $\operatorname{EGAR}(m)$ & Homoschedastic GAR components & $T + \phi$ \\
\quad  $\operatorname{EGARI}(m)$ & EGAR${}+{}$isotropic for $D$ components &$ 1 + \phi$\\
\hline
\end{tabular*}
\tabnotetext[*]{t1}{$\phi$ refers to the number of parameters of the generic
$\Phi_i$: $\phi=\frac{T(T-1)}{2}-\frac{(T-m-1)(T-m)}{2}$. }
\end{table}

With reference to the temporal ``between'' covariance matrices,
$\Phi_i$, besides the heteroscedastic situation for different
values of $m$, we also model the scenarios of homoscedastic
components $\Phi_i=\Phi$ for all $i$, and of isotropic constraint
$D_i=d_i I_T$, which implies that all the innovation parameters do
not depend on the time, thus modeling an autoregressive process.

With regard to the ``within'' covariance matrix $\Omega_i$, we
consider the spectral decomposition parameterization given in
\citet{Celeux1995} and \citet{Banfield1993} and used by
\citet{Viroli2011} in
mixtures of matrix-normal distributions. This parameterization
consists in expressing\vspace*{1pt} $\Omega_i$ in terms of its eigenvalue
decomposition as $\Omega_i=\lambda_iV_i{A}_i{V}^\top_i$, where
${V}^\top_i$ is the matrix of eigenvectors, ${A}_i$ is a diagonal
matrix whose elements are proportional to the eigenvalues of
$\Omega_i$ and $\lambda_i$ is the associated constant of
proportionality. By considering homoschedastic or varying
quantities across the mixture components, different submodels can
be defined using the nomenclature in \citet{Fraley2002}: VVV
refers to heteroscedastic components with respect to the within
covariance matrix, EEE indicates components with homoscedastic
within covariance matrices, VVI denotes diagonal but varying
variability components, EEI refers to diagonal and homoscedastic
components and, finally, VII and EII denote spherical components
with and without varying volume. For an exhaustive summary of the
covariance pattern structures see Table~\ref{tabcovariancestructure}.

Therefore, a large family of possible mixture models can be
defined, allowing for special pattern structures on both the temporal
and response covariance
matrices. In this family, the model
parameters can be efficiently estimated through the EM algorithm
which alternates between the expectation and the maximization
steps until convergence. Model selection can be performed by
the BIC and AIC information criteria. In the next section model
fitting is developed and illustrated.

\subsection{Model validation}\label{sec34}
In order to validate the model and to explore its robustness and its
accuracy, several simulation studies were performed.
The results show that the model is robust in finding the true temporal
structure (when it actually exists); furthermore, it yields a good
classification of the units and a good estimate accuracy even when the
model was misspecified. Finally, the algorithm recovered the true
number of groups in the majority of the cases, regardless of the
correct specification of the temporal structure.

A full description is given in the Supplementary Material [\citet{suppA}].

\section{Likelihood inference}\label{sec4}

The model parameters can be efficiently estimated through the EM
algorithm, where the missing data are the group membership labels
[\citet{Dempster1997}]. Let $z_j$ be the vector of dimension $k$
denoting the component membership of each matrix sample, $Y_j$.
Then the complete-data likelihood of the proposed pattern mixture
model is given by
%
\begin{eqnarray}
L_c(Y,z;\bolds{\pi},\bolds{\Theta})&=&\prod
_{j=1}^n\prod_{i=1}^k
f(Y_j,z_{ji};\bolds{\pi},\bolds{\Theta})
\nonumber
\\[-8pt]
\label{eqn3}
\\[-8pt]
\nonumber
&=&\prod_{j=1}^n\prod
_{i=1}^k \bigl(\pi_i
\phi^{(T \times
p)}(Y_j;M_{ij},\Phi_i,
\Omega_i) \bigr)^{z_{ji}},
\end{eqnarray}
where $\bolds{\pi}=\{\pi_1,\ldots,\pi_k\}$ and
$\bolds{\Theta}=\{\bolds{\Theta}_1,\ldots,\bolds{\Theta}_k\}$.

Given the allocation variable, the complete density
$f(Y,z;\bolds{\pi},\bolds{\Theta})$ defined in (\ref{eqn3})
can be decomposed into the product of the two densities
\begin{eqnarray*}
&& f(z|\bolds{\pi},\bolds{\Theta})= \prod_{i=1}^k
\pi_i^{z_i},
\end{eqnarray*}
from which $f(z_i=1|\bolds{\pi},\bolds{\Theta})= \pi_i$ and
$f(Y|z_i=1;\bolds{\pi},\bolds{\Theta})=\phi^{(T \times
p)}(Y;M_{ij}, \Phi_i,\Omega_i)$.

The conditional expectation of the complete log-density given the
observable data, using a fixed set of parameters $\bolds{\pi}'$
and $\bolds{\Theta}'$, is
%
\begin{eqnarray}
&& \arg\max_{\bolds{\pi},\bolds{\Theta}}E_{z|Y;
\bolds{\pi}',\bolds{\Theta}'} \bigl[ \log f (Y,z|\bolds{
\pi},\bolds{\Theta} ) \bigr]
\nonumber
\\[-8pt]
\label{eqn14}
\\[-8pt]
\nonumber
&&\qquad =\arg\max_{\bolds{\pi},\bolds{\Theta}}E_{z|Y;
\bolds{\pi}',\bolds{\Theta}'} \bigl[ \log f (Y|z;
\bolds{\pi},\bolds{\Theta} ) + \log f (z|\bolds{\pi},\bolds{\Theta} ) \bigr],
\end{eqnarray}
which is equivalent to maximizing the following function with
respect to $\bolds{\pi}$ and~$\bolds{\Theta}$:
\begin{eqnarray*}
&& Q (\bolds{\pi},\bolds{\Theta}| \mathbf{Y},\bolds{\tau} )\\
&&\qquad=\sum
_{i=1}^k \sum_{j=1}^n
\tau_{ij} \log \bigl[ \pi_i \phi^{(T \times
p)}(Y_j;M_{ij},
\Phi_i,\Omega_i) \bigr]
\\
&&\qquad= \sum_{i=1}^k n_i \log
\pi_i - \frac{Tpn}{2}\log2\pi- \sum_{i=1}^k
\sum_{j=1}^n \tau_{ij}
\log|D_i|^{-{p}/{2}}\\
&&\qquad\quad {}- \sum_{i=1}^k
\sum_{j=1}^n \tau_{ij}\log|
\Omega_i|^{-{T}/{2}}
\\
&&\qquad\quad{}- \frac{1}{2}\sum_{i=1}^k\sum
_{j=1}^n \tau_{ij}
\operatorname{tr} \bigl\{ \bigl(U_i^\top
D_i^{-1}U_i\bigr) (Y_j-\tilde
X_j\tilde\Theta_i)\Omega_i
^{-1}(Y_j-\tilde X_j\tilde
\Theta_i)^\top \bigr\}
\\
&&\qquad= \sum_{i=1}^k n_i \log
\pi_i - \frac{Tpn}{2}\log2\pi- \sum_{i=1}^kn_i
\frac{p}{2}\log|D_i| - \sum_{i=1}^kn_i
\frac{T}{2} \log|\Omega_i|
\\
&&\qquad\quad{}- \sum_{i=1}^k \frac{n_i}{2}
\operatorname{tr} \bigl\{ \bigl(U_i^\top
D_i^{-1}U_i\bigr) S_i \bigr\},
\end{eqnarray*}
where $\mathbf{Y}=Y_1,\ldots,Y_n$, $n_i=\sum_{j=1}^n\!\tau_{ij}$,
$S_i=(1/n_i)\sum_{j=1}^n\!
\tau_{ij}(Y_j-\tilde X_j\tilde\Theta_i)\Omega_i ^{-1}(Y_j-\tilde
X_j\tilde\Theta_i)^\top$
and $\bolds{\tau}=\{\tau_{ij}\}$ are the posterior probabilities
$f(z_{ij}|Y_j;
\bolds{\pi},\bolds{\Theta})$ derived for a fixed set of
parameters by the Bayes's theorem
[\citet{McLachlan2000}] as
%
\begin{equation}
\tau_{ij} =\frac{\pi_i\phi^{(T \times p)}(Y_j;M_{ij},\Phi_i,\Omega_i)}{
\sum_{h=1}^k\pi_h \phi^{(T \times
p)}(Y_j;M_{hj},\Phi_h,\Omega_h)}.
\end{equation}

By maximizing (\ref{eqn14}) the parameter estimates for given
values of $m$ and $k$ and a fixed pattern structure can be obtained.
All the estimates are in closed form. With reference to the weights, we
have $\hat{\pi}_i=\frac{\sum_{j=1}^n\tau_{ij}}{n}$.

The estimators of the regression coefficients are
\begin{eqnarray*}
\hat{\tilde\Theta}_i&=& \Biggl[ \sum_{j=1}^n
\tau_{ij}\bigl(\tilde X_j^\top
\Phi_i^{-1} \tilde X_j\bigr)
\Biggr]^{-1}{\sum_{j=1}^n
\tau_{ij}\tilde X_j^\top\Phi_i^{-1}Y_j}.
\end{eqnarray*}
With reference to the temporal covariance matrices, the derivative of
(\ref{eqn14}) with respect to $D_i$ leads to $\hat{D}_i=\frac
{U_iS_iU_i^\top}{p}$.
The estimates of the parameters contained in $U_i$ can be obtained by solving
%
\begin{eqnarray}
\label{sistem} && \frac{\partial Q  (\bolds{\pi},\bolds{\Theta}|
\mathbf{Y},\bolds{\tau} )}{\partial U_i} = -n_i D_i^{-1}U_iS_i=0.
\end{eqnarray}
Since only the  lower triangular
part of $U_i$ contains the autocorrelations to be estimated, the
expression in (\ref{sistem}) leads to a system of simple linear
equations for each $r=2,\ldots,T$ that have the closed-form solution
\begin{eqnarray*}
&& \pmatrix{ %
\hat{\rho}_{r,r-m}
\cr
\hat{\rho}_{r,r-m+1}
\cr
\cdots
\cr
\hat{\rho}_{r,r-1} } \\
&&\qquad=\pmatrix{ s_{r-m,r-m} &
s_{r-m+1,r-m} & \cdots & s_{r-1,r-m}
\cr
s_{r-m,r-m+1} &
s_{r-m+1,r-m+1} & \cdots & s_{r-1,r-m+1}
\cr
\cdots & \cdots & \cdots &
\cdots
\cr
s_{r-m,r-1} & s_{r-m+1,r-1} & \cdots & s_{r-1,r-1}
}^{-1} \pmatrix{ s_{r,r-m}
\cr
s_{r,r-m+1}
\cr
\cdots
\cr
s_{r,r-1} },
\end{eqnarray*}
where $r=2,\ldots,T$ and $s$ are the elements of $S_i$.

Finally, the estimator of the pattern structure of the within
covariance matrices under the general form VVV is
\begin{eqnarray*}
\hat{\Omega}_i&=&\frac{\sum_{j=1}^n\tau_{ij}(Y_j-\tilde X_j
\tilde\Theta_i)^\top\hat{\Phi}_i^{-1}(Y_j-\tilde X_j
\tilde\Theta_i)}{T\sum_{j=1}^n\tau_{ij}}.
\end{eqnarray*}

The solution is unique up to a multiplicative constant, say $a\neq0$,
since $\Phi_i \otimes\Omega_i= a\Phi_i \otimes\frac{1}{a}\Omega_i$.
In practice, a way to obtain a unique solution is to impose the
identifiability constraint $\operatorname{tr}\Omega_i=p$ or, alternatively, $\sum_{h,c}\omega_{ihc}=p^2$, where $h$ and $c$ indicate the rows and
columns of $\Omega_i$ and $\omega$ is the single element of $\Omega_i$.

The estimator under the other parameterizations can be obtained in a
similar way [see \citet{Viroli2011} and \citet{McNicholas2010}].

Once the maximum likelihood estimates have been obtained, the standard
errors of the regression coefficients may be computed in order to
identify the significant covariates in each group of subjects. These
may be obtained on the basis of the observed information matrix,
$\mathcal{I_H}(\hat{\bolds{\Theta}})=-\sum_{j=1}^n[\mathbf{Q}_j(\hat{\bolds{\Theta}})]$, where $\mathbf{Q}_j$ is the Hessian matrix of
the likelihood function evaluated at its maximum for observation $j$
with $j=1,\ldots,n$, computed using the package \texttt{numDeriv} of
\texttt{R}. The algorithm has been implemented in R code and it is available
upon request.

\section{Case study: HRS panel data} \label{sec5}

In order to adequately model the data, we estimated the proposed CPMM,
both with and without the inclusion of covariates, allowing for a
different number of components (i.e., $k=1,\ldots, 5$), for
different structures for~$\Omega$ (the nontemporal patterns in Table~\ref{tabcovariancestructure}), for different structures for~$\Phi$
(i.e., GAR, GARI, EGAR, EGARI and all the
nontemporal structures), and for different
orders of the generalized autoregressive process (i.e., for
$m=0,\ldots, T-1=5$, where $m=0$ indicates time-independent data).
All of these models have been estimated in a multistart strategy,
so as to avoid possible EM problems of local maxima.

\begin{table}[b]
\caption{Estimation results of HRS data of the proposed CPMM model and
of the multivariate mixed model CPMM with random time-specific
intercept (CPMM-i) and random slopes (CPMM-is). The table shows the
computational time (in seconds), the maximum log-likelihood, the
Bayesian Information Criterion (BIC), the preferred number of clusters
$k^*$ and the preferred structure of the two CPMM covariance matrices
according to the BIC. In the last column, the root-mean-square
deviation (RMSD) of the predicted values by the fitted models is
reported}\label{tabexresults}
\begin{tabular*}{\tablewidth}{@{\extracolsep{\fill}}lccccccc@{}}
\hline
 & \textbf{Time} &  &  & & &  & \\
\textbf{Model}& \textbf{(sec.)} & \textbf{logLik}&
\textbf{BIC} &  $\bolds{k^*}$ & $\bolds{\Phi}$ & $\bolds{\Omega}$& \textbf{RMSD} \\
\hline
CPMM (no cov.)& \phantom{0}26 & $-$14{,}129 & 28{,}940 & 4 & EGAR $m=4$ & VVV & 2.53 \\
CPMM (with cov.) & \phantom{0}13 & $-$14{,}030 & \textbf{28{,}777} & 3 & EGAR $m=3$ & VVV
& 2.42\\
LCMM-i (no cov.) & \phantom{00}7 & $-$15{,}184 & 30{,}439 & 1 & -- & -- & 3.16\\
LCMM-i (with cov.)& 267 & $-$14{,}802 & 29{,}780 & 2 & -- & -- & 2.95\\
LCMM-is (with cov.) & 425 & $-$14{,}801 & 29{,}802 & 2 & -- & -- & 2.95\\
\hline
\end{tabular*}
\end{table}

For comparative purposes, we have also estimated latent class mixed
models for longitudinal data with the R package \texttt{lcmm},
allowing for models with a different
number of clusters (i.e., $k=1,\ldots, 5$). We considered the models
with and without covariates, with random intercepts only and with both
random intercepts and slopes. On the HRS data we have also tried to
estimate the growth mixtures models with \texttt{Mplus};
unfortunately, we have encountered convergence problems of the
algorithm with
$k>1$.

A summary of the estimated models is reported in Table~\ref{tabexresults}, where the best number of clusters, $k^*$, for each family
of models according to the Bayesian Information Criterion (BIC) is shown.
The table shows the computational time (in seconds), the maximum
log-likelihood, the value of the BIC, the preferred number of clusters
$k^*$ by BIC and the preferred structure of the two CPMM covariance matrices.

In order to compare the adequacy of the different fitted models, we
have also computed a predictive measure of their performance, given by
the root-mean-square deviation (RMSD) of the predicted values:
\[
\mathrm{RMSD}=\sqrt{\frac{\sum_{j=1}^n\sum_{t=1}^T\sum_{h=1}^p (\hat
{y}_{jth}-{y}_{jth})^2}{T\cdot n\cdot p}}.
\]

The best fit of the data according to the BIC is the one obtained by
the CPMM model with the inclusion of covariates, that consists of
$k=3$ components; they are heteroscedastic with respect to the
within covariance matrix $\Omega$ (i.e., structure ``VVV''), and they have
a ``EGAR'' structure with $m=3$ for the temporal covariance
matrix $\Phi$. The second preferred model is again the
CPMM, but without the inclusion of covariates; this modeling
requires a further component in order to explain heterogeneity in
the data and a larger autoregressive order. The same insight is given
by the RMSDs that are a measure of the adequacy of the fitted models in
terms of their predictive performance.

The latent class mixed model with no covariates fails to find a
clustered structure; when
covariates are included, the preferred model consists in two classes,
but there is no
specification of the temporal pattern and the predictive performances
are worse.

\begin{table}[b]
\tablewidth=200pt
\caption{HRS data: group sizes ($n_i$) and means of the observed
variables separately by group}\label{tabexsummary}
\begin{tabular*}{200pt}{@{\extracolsep{\fill}}lccc@{}}
\hline
& \textbf{Group 1} & \textbf{Group 2} & \textbf{Group 3}\\
\hline
$n_i$ & 60\phantom{.00} & 187\phantom{.00} & 112\phantom{.00} \\[3pt]
Episodic memory & \phantom{0}5.11 & \phantom{00}7.59 & \phantom{00}9.08 \\
Mental status & \phantom{0}7.89 & \phantom{00}9.99 &\phantom{0}11.18 \\
Mood & \phantom{0}8.77 & \phantom{00}8.46 & \phantom{0}11.38 \\
\hline
\end{tabular*}
\end{table}

\begin{figure}

\includegraphics{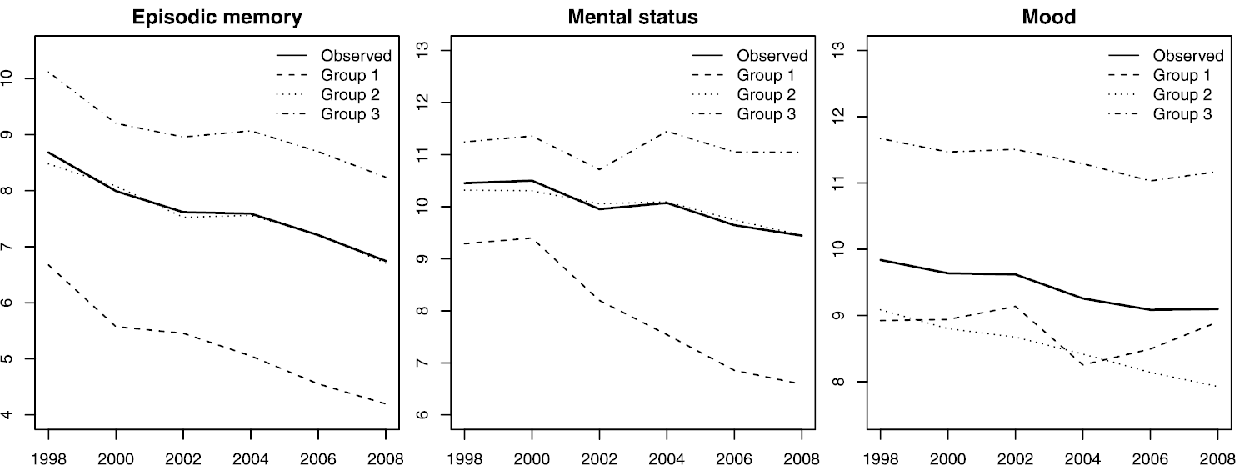}

\caption{Mean profiles of the three responses according to cluster
memberships. The black solid lines represent the empirical mean
values.}\vspace*{20pt}\label{figclmeanprofiles}
%

\includegraphics{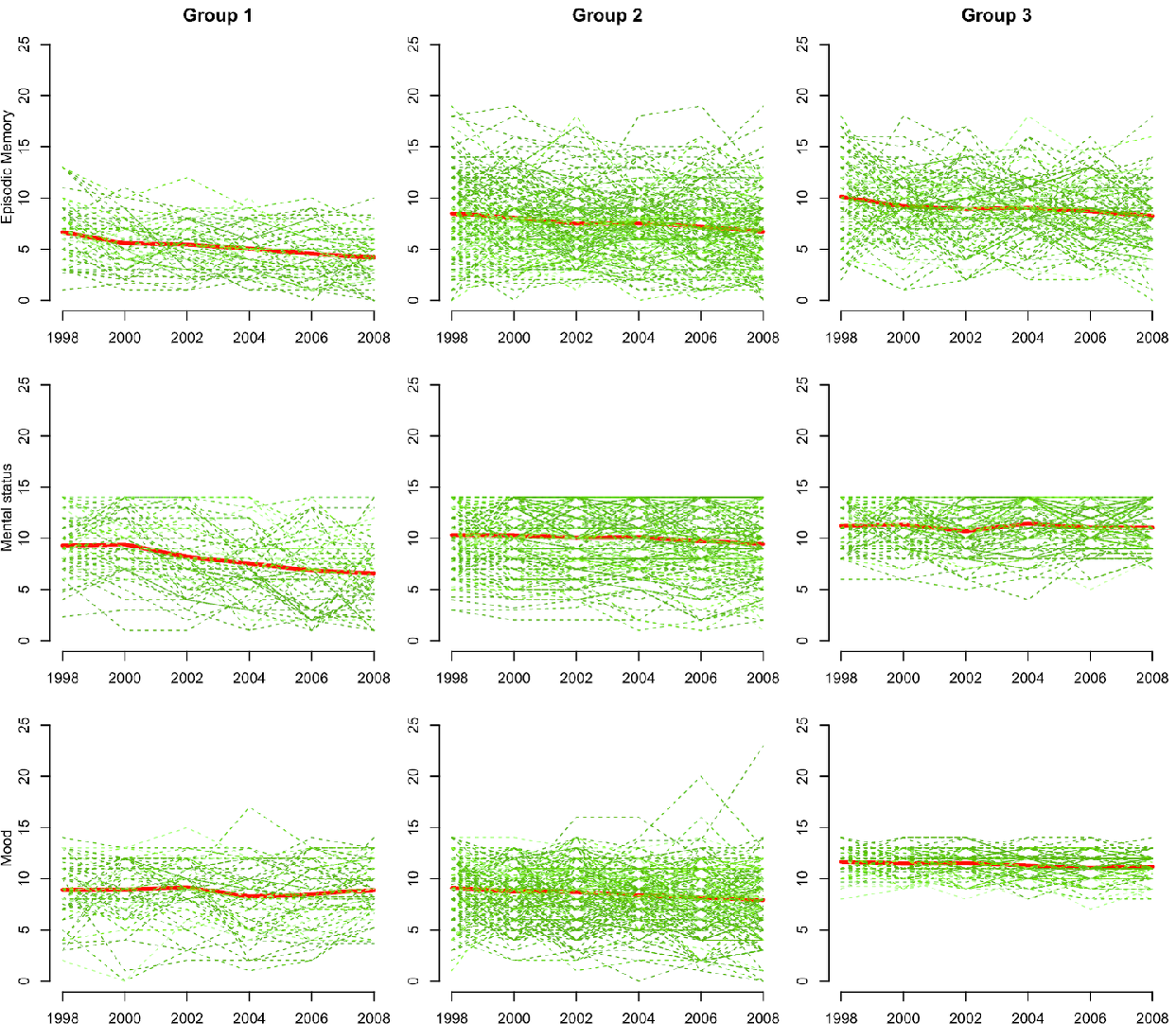}

\caption{Responses of profiles according to cluster
memberships. The solid red lines represent the predicted
class-conditional mean profiles. Individual trajectories (dashed lines)
have been color coded such that more intense colors correspond to
higher posterior probabilities.}\label{figwithinclusters}
\end{figure}

The three groups of the selected CPMM model consist of 60, 187 and 112
individuals, respectively. Table~\ref{tabexsummary} summarizes
the mean values of the three responses in the obtained clusters. Groups
look easily interpretable. In fact, by looking at the mean values in
Table~\ref{tabexsummary}, people in Group 1 are those with the
lowest episodic memory and mental status, yielding to a
moderate low mood; respondents in Group~3 are on average
the happiest, those with the highest score in mental status and
episodic memory. Finally, individuals in Group 2 place in an
intermediate position with respect to the others.

%
\begin{table}[t]
\tabcolsep=0pt
\caption{Regression coefficient estimates and standard errors of the
model with $k=3$ groups, separately for each response (EM${}={}$Episodic
Memory, MS${}={}$Mental Status and MO${}={}$Mood). The $p$-values are referred to
as the asymptotic Wald test}
\label{tabexsignif}
\begin{tabular*}{\tablewidth}{@{\extracolsep{\fill}}llccccccccc@{}}
\hline
& & \multicolumn{3}{c}{\textbf{Group 1}} & \multicolumn{3}{c}{\textbf{Group 2}} &
\multicolumn{3}{c@{}}{\textbf{Group 3}} \\[-4pt]
& & \multicolumn{3}{c}{\hrulefill} & \multicolumn{3}{c}{\hrulefill}& \multicolumn{3}{c@{}}{\hrulefill}\\
& & \textbf{EM} & \textbf{MS} & \textbf{MO} & \textbf{EM} & \textbf{MS} & \textbf{MO} &
\textbf{EM} & \textbf{MS} & \textbf{MO} \\
\hline
Gender  & Estimates & $-$0.96\phantom{0} & \textbf{$\bolds{-}$2.71}\phantom{0} & \phantom{$-$}0.65\phantom{0} &
\phantom{$-$}\textbf{1.67}\phantom{0} & \phantom{$-$}0.19\phantom{0} & \textbf{$\bolds{-}$0.93}\phantom{0} & \phantom{$-$}\textbf{1.42}\phantom{0} & \phantom{$-$}0.25\phantom{0} &
$-$0.07\phantom{0} \\
& St.Err. & \phantom{$-$}0.90\phantom{0} & \phantom{$-$}1.25\phantom{0} & \phantom{$-$}0.96\phantom{0} &
\phantom{$-$}0.45\phantom{0} & \phantom{$-$}0.25\phantom{0} & \phantom{$-$}0.47\phantom{0} & \phantom{$-$}0.71\phantom{0} & \phantom{$-$}0.54\phantom{0} &
\phantom{$-$}0.26\phantom{0}\\
& $p$-value & \phantom{$-$}0.144 & \phantom{$-$}0.015 & \phantom{$-$}0.249 & \phantom{$-$}0.000 & \phantom{$-$}0.224 & \phantom{$-$}0.024 & \phantom{$-$}0.022 &
\phantom{$-$}0.320 & \phantom{$-$}0.389\\[3pt]
Age & Estimate & \textbf{$\bolds{-}$0.07}\phantom{0} & \phantom{$-$}0.01\phantom{$-$} & $-$0.00\phantom{0} &
\textbf{$\bolds{-}$0.16}\phantom{0} & \textbf{$\bolds{-}$0.05}\phantom{0} & \phantom{$-$}0.05\phantom{0} & \textbf{$\bolds{-}$0.08}\phantom{0} & \phantom{$-$}0.01\phantom{0} &
$-$0.00\phantom{0}
\\
& St.Err. & \phantom{$-$}0.02\phantom{0} & \phantom{$-$}0.02\phantom{0} & \phantom{$-$}0.03\phantom{0} &
\phantom{$-$}0.04\phantom{0} & \phantom{$-$}0.03\phantom{0} & \phantom{$-$}0.04\phantom{0} & \phantom{$-$}0.02\phantom{0} & \phantom{$-$}0.04\phantom{0} &
\phantom{$-$}0.02\phantom{0}\\
& $p$-value & \phantom{$-$}0.001 & \phantom{$-$}0.329 & \phantom{$-$}0.454 & \phantom{$-$}0.000 & \phantom{$-$}0.015 & \phantom{$-$}0.078 & \phantom{$-$}0.001 &
\phantom{$-$}0.406 & \phantom{$-$}0.492\\[3pt]
Education  & Estimate & \phantom{$-$}0.05\phantom{0} & \phantom{$-$}0.05\phantom{0} & \phantom{$-$}0.02\phantom{0} &
\phantom{$-$}\textbf{0.41}\phantom{0} & \phantom{$-$}\textbf{0.48}\phantom{0} & \phantom{$-$}\textbf{0.11}\phantom{0} &
\phantom{$-$}\textbf{0.24}\phantom{0} & \phantom{$-$}\textbf{0.25}\phantom{0}
& \phantom{$-$}0.05\phantom{0} \\
& St.Err. & \phantom{$-$}0.11\phantom{0} & \phantom{$-$}0.10\phantom{0} & \phantom{$-$}0.12\phantom{0} & \phantom{$-$}0.06\phantom{0} &
\phantom{$-$}0.03\phantom{0} & \phantom{$-$}0.05\phantom{0} & \phantom{$-$}0.09\phantom{0} & \phantom{$-$}0.08\phantom{0} &
\phantom{$-$}0.04\phantom{0}\\
& $p$-value & \phantom{$-$}0.335 & \phantom{$-$}0.331 & \phantom{$-$}0.450 & \phantom{$-$}0.000 & \phantom{$-$}0.000 & \phantom{$-$}0.010 & \phantom{$-$}0.002 &
\phantom{$-$}0.001 & \phantom{$-$}0.101\\[3pt]
Health & Estimate & \phantom{$-$}0.09\phantom{0} & \phantom{$-$}0.05\phantom{0} & \textbf{$\bolds{-}$0.67}\phantom{0} & $-$0.09\phantom{0} & $-$0.05\phantom{0} &
\textbf{$\bolds{-}$0.62}\phantom{0} & $-$0.15\phantom{0} & $-$0.04\phantom{0} &\textbf{$\bolds{-}$0.33}\phantom{0} \\
self-rating & St.Err. & \phantom{$-$}0.24\phantom{0} & \phantom{$-$}0.32\phantom{0} & \phantom{$-$}0.40\phantom{0} & \phantom{$-$}0.13\phantom{0}
& \phantom{$-$}0.09\phantom{0} & \phantom{$-$}0.11\phantom{0} & \phantom{$-$}0.22\phantom{0}
& \phantom{$-$}0.18\phantom{0} & \phantom{$-$}0.08\phantom{0}\\
& $p$-value & \phantom{$-$}0.355 & \phantom{$-$}0.435 & \phantom{$-$}0.045 & \phantom{$-$}0.249 & \phantom{$-$}0.302 & \phantom{$-$}0.000 & \phantom{$-$}0.253 &
\phantom{$-$}0.405 & \phantom{$-$}0.000\\
\hline
\end{tabular*}
\end{table}

Figure~\ref{figclmeanprofiles} gives a visual representation of
the cluster means for each response along time; the observed mean
profile is in between the mean profiles of subjects in Groups 1 and 3,
partially overlapping the profile of Group 2 members.

\begin{figure}[t]

\includegraphics{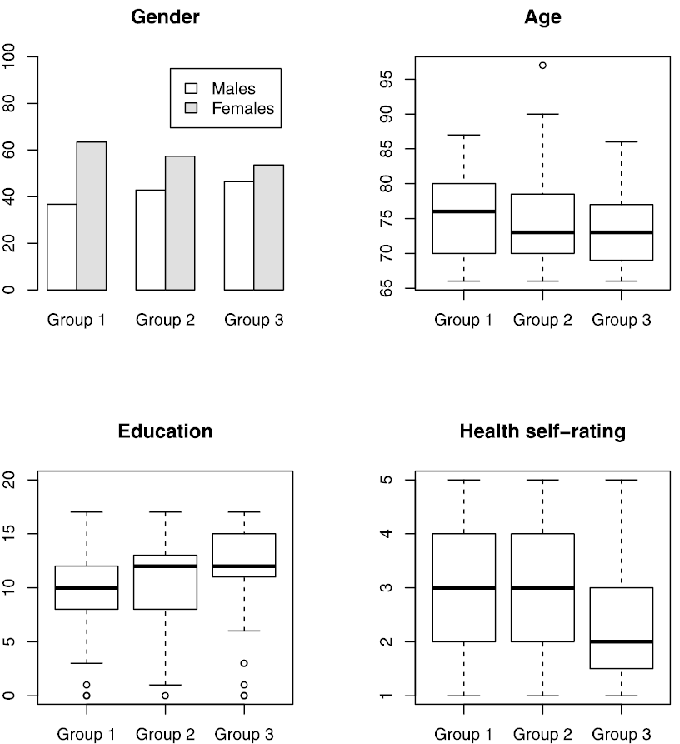}

\caption{Covariate distributions conditional on groups. The panel \textup{(a)}
shows the histograms of the males and females separately for each
group. Panels \textup{(b)}, \textup{(c)} and \textup{(d)} show the boxplots of the distribution of
the covariates ``Age,'' ``Education'' and ``Health self-rating''
conditional on the three groups.}\vspace*{-4pt}\label{figclcovariates}
\end{figure}

Following \citet{Erosheva2014}, we also plotted predicted group
mean trajectories for each response (solid red lines) along with the
observed trajectories (dashed green lines) from the individual
classified in each group (Figure~\ref{figwithinclusters}). The
individual trajectories have been color coded such that the more
intense green corresponds to higher posterior probability. These
graphics allow us to visualize how much of the individual variability
is explained by group means, how much the identified groups overlap and
how stable the classification is.
It is evident that there is not a clear separation between individual
trajectories observations from different groups, although\vadjust{\goodbreak} the
classification is quite stable since the posterior probabilities for
most individuals are close to 1.

Furthermore, our approach allows to estimate regression coefficients
separately on groups and $p$-values are computed according to the Wald
test to check significance. Table~\ref{tabexsignif} contains the
regression coefficient estimates (significant values are denoted in
bold). The interesting point is that covariates may or may not have a
significant effect on some responses depending on groups; the
contribution of each regressor on the dependent variables
according to group membership is a free benefit of our proposed
model. Indeed, as an example consider the variable ``Education.''
It has a significant positive effect on ``Episodic memory'' and on
``Mental status'' as far as respondents belong to Group 2 or 3;
therefore, it may mean that for people in Group 1 which are on
average older and less educated, one year more of education would not
determine any significant change in any of the responses, whereas
it may improve the mood of people with features similar to Group 2
members. Conversely, the ``Self-rating health'' has significant negative
effect on the ``Mood'' (remember that this response has a reverse
scale), independently on group membership; the same global
negative effect is carried out by age on the episodic memory.

\begin{table}[b]
\tablewidth=220pt
\caption{Estimated temporal correlation matrix}\label{tabestphi}
\begin{tabular*}{220pt}{@{\extracolsep{\fill}}lcccccc@{}}
\hline
&\textbf{1998} & \textbf{2000} & \textbf{2002} & \textbf{2004} & \textbf{2006} & \textbf{2008} \\
\hline
1998 & 1.000 & & & & &\\
2000 & 0.408&1.000 & & & & \\
2002 &0.407 &0.449 &1.000 & & & \\
2004 &0.345 &0.378 &0.487 & 1.000 & & \\
2006 &0.260 &0.355 &0.455 &0.481 & 1.000 & \\
2008 &0.225 &0.272 &0.415 &0.448 &0.525 &1.000\\
\hline
\end{tabular*}
\end{table}

\begin{table}[t]
\tablewidth=200pt
\caption{Estimated responses correlation matrix across the three
groups}\label{tabestomega}
\begin{tabular*}{200pt}{@{\extracolsep{\fill}}lccc@{}}
\hline
& \textbf{Episodic} & \textbf{Mental} &  \\
& \textbf{memory} & \textbf{status} & \textbf{Mood}\\
\hline
Episodic memory & 1.00\phantom{0} & & \\
Mental status & 0.155 & 1.000 & \\
Mood & 0.020 & 0.005 & 1.000\\
\hline
\end{tabular*}
\end{table}

Some further insight can be also offered by the covariate distributions
conditional on groups, so that differences in the attributes can
be highlighted. From Figure~\ref{figclcovariates} we can see that
Group 1 has
the highest prevalence rate of females with respect to males; its
respondents are older than\vadjust{\goodbreak} individuals in other groups and look
less educated. Remembering that the health self-rating variable
has a reverse scale, respondents in Group 3 scored lower points
than individuals in Group 1. Indeed, people in the former group
are the youngest and the most educated with respect to the whole
sample. This characterization is consistent with the response mean
values.

Finally, Tables~\ref{tabestphi} and \ref{tabestomega} contain the
estimated temporal and responses correlation matrices, respectively.
They refer to the error term and if compared to the observed ones (see
Tables~\ref{tabcortemp} and \ref{tabcorp}), we can see that
estimated entries are smaller. This means that the introduction of the
covariates into the model actually explained a large part of the
observed correlation.\vadjust{\goodbreak}

\section{Discussion}\label{sec6}
The results presented in Section~\ref{sec5} allow for an accurate
description of the cognitive functioning in elderly people, by allowing
for an identification of three groups of individuals that share a
common response pattern. The partial overlap of the groups may suggest
that this is an approximation to an underlying continuum of variability
in different temporal patterns.

In particular, it is possible to identify a group of respondents (Group
3) that scored on average the highest results on the tests and, hence,
those that require comparatively less attention. These individuals are
on average younger and could benefit from more years of education, as
it has a significant impact on episodic memory and mental status.
Females in this group have an advantage in episodic memory compared to
men. As one may expect, age has a negative effect on memory, but it is
less remarkable compared to other groups.

Conversely, members of Group 1 are approximately the oldest in age and
those who received less years of education. This is the more
problematic set, since individuals are more depressed and they reported
the lowest scores in episodic memory and mental status. For these
respondents, their perceived health status is an important determinant
of their mood, whereas education does not significantly affect any of
their responses. Females in this group have a large disadvantage in
mental status: this result tells us that interventions should target
elderly ladies by developing psychiatric and health assistance.

Finally, the last considered group is the one whose members obtained
scores that lie in between the two extremes (Group 2). Mood is
significantly affected by the respondents perceived health status and
it is significantly worse for females, on average. Age has an important
effect on episodic memory and on mental status, but those negative
effects are balanced by education: this is an important determinant for
all the responses.
Efforts here should be directed to improving heath assistance and by
creating or reinforcing moments of social aggregation where cultural
initiatives are promoted. Particularly, attention should be focused on
individuals that received less education: these individuals had lower
performances in all responses. Members of this group (and later members
of Group 3) could be the future Group~1, so it is important to prevent
and to dedicate care to the aspects that would be more crucial in the future.

Since the elderly ladies resulted in having more mental issues, some
awareness campaigns to sensitize public opinion on female care and
assistance should be promoted; the same philosophy should guide funds
allocation decisions and tax reduction policies, particularly when
dealing with associations that serve health and mental assistance.

Cognitive impairment and depression are, in fact, costly. States should
consider developing a comprehensive action plan to respond to the needs
of people with cognitive impairment and their families, to empower
people to seek help and to support recovery, involving different
agencies, as well as private and public organizations, and to expand
research on this topic. Further investigations on other possible
determinants of the cognitive functioning (such as genetic
predisposition and presence of important comorbidities) can be
explored, so as to highlight other features of the phenomenon and to
better understand its temporal evolution.

\section{Concluding remarks}\label{sec7}
In this work we have presented a novel approach for modeling
multivariate longitudinal data in the presence of unobserved heterogeneity.
It is defined as a particular linear mixed model with discrete
individual random intercepts, but differently from the standard random
effects models, the proposed CPMM does not require the usual local
independence assumption; in this way the temporal structure and the
association among the responses can be explicitly modeled.

The proposal has the benefit of being very flexible and parsimonious at
the same time, provided that specific pattern structures are suitably
chosen in the model selection phase. Its flexibility freely adds
meaningful interpretation to the study under analysis since, besides
the temporal dependence and the response association (that can be both
class-specific), it allows for a different contribution of each
regressor on the responses according to group membership. In so doing,
the identified groups receive a global and accurate phenomenal
characterization, as shown in the HRS application. From the
computational point of view, the algorithm is pretty fast (in our real
application a few seconds are required) compared
to the alternative approaches, and no convergence problems have been observed.

The price to be paid for this great flexibility and computational
feasibility is connected to the kind of data structures that can be
analyzed when the matrix-normal distribution is assumed: this
probabilistic model implies that the observed times and the number of
responses are equally spaced and balanced. This aspect could limit the
applicability of the proposed approach to all the observational studies
where the number of responses is not constant across times and subjects
or when data are incomplete. On the other hand, the extension of the
model to deal with incomplete data under the missing at random (MAR)
mechanism could be developed in the same framework of the EM algorithm
by splitting each set of observations into the missing and observed
components through permutation matrices. This issue and the related
estimation scheme are aspects that deserve further attention.
Furthermore, in our formulation we confined our attention to continuous
responses. A natural extension consists of generalizing our model
to either binary or categorical response variables (or mixed-type).
This extension may be performed by considering generalized
matrix-regression models with discrete random intercepts, although new
computational problems would be involved.\looseness=-1

\section*{Acknowledgments}
We are grateful to the Center for the Study
of Aging of US (\surl{http://www.rand.org}) for the data use agreement.

\begin{supplement}[id=suppA]
\stitle{Simulation study}
\slink[doi]{10.1214/15-AOAS816SUPP} 
\sdatatype{.pdf}
\sfilename{aoas816\_supp.pdf}
\sdescription{The supplementary material contains the description and
the results of the simulation studies that involved and investigated
many aspects of the model validation.}
\end{supplement}


\printaddresses
\end{document}